# Paramagnetic $Fe_xTa_{1-x}$ alloys for engineering of perpendicularly magnetized tunnel junctions

Matthias Gottwald[1], Jimmy J. Kan[1], Kangho Lee[2], Seung H. Kang[2], and [*]Eric E. Fullerton[1]

[1]Center for Magnetic Recording Research, University of California-San Diego, La Jolla, CA 92093-0401, USA

[2]Advanced Technology, Qualcomm Incorporated, San Diego, CA 92121-1714, USA

**Abstract:** Exchange coupling between two magnetic layers through an interlayer is of broad interest for numerous recent applications of nano-magnetic systems. In this Letter we study ferromagnetic exchange coupling through amorphous paramagnetic Fe-Ta alloys. We show that the exchange coupling depends exponentially on spacer thickness and scales with the Fe-Ta susceptibility, which can be tuned via the alloy composition and/or temperature. Such materials are of high interest for the engineering of perpendicularly magnetized CoFeB-MgO based tunnel junctions as it enables ferromagnetic coupling of magnetic layers with differing crystalline lattices, suppresses dead layers and can act as an inter-diffusion barrier during annealing.

[*]Author to whom correspondence should be addressed. Email: efullerton@ucsd.edu

I. Introduction

Controlled magnetic interlayer coupling has enabled a broad range of new magnetic nanotechnologies. The study of antiferromagnetic (AF) coupling was closely related to the discovery of giant magnetoresistance[1,2]. Synthetic antiferromagnets(SAF), which are made up of two AF-coupled ferromagnetic layers, are used as the reference layers in magnetoresistance field sensors in hard disk drives [3] and magneto-resistance random access memories (MRAM) [4]. AF-coupled layers are also used for free layers in field-written MRAM devices [5], as longitudinal recording media [6-8] and as soft underlayers for perpendicular magnetic recording media [9]. More recently ferromagnetic interlayer coupling has emerged as a way to tune the properties of magnetic devices. Ferromagnetic coupling provides a pathway to engineer incoherent magnetic reversal that allows a reduction of the magnetic field required to write magnetic recording media and further allows control of inter-granular exchange [8, 10-12]. A similar approach has been proposed to lower the write current in spin-transfer-torque (STT) MRAM devices [13] and to tune the operating frequencies of planar microwave [14] and spin-torque devices [15].

In this letter we detail the strong and tunable ferromagnetic interlayer coupling across amorphous paramagnetic Fe-Ta alloys. For Fe contents less than 90 at.% Fe-Ta alloys are amorphous and paramagnetic at room temperature [16]. Such materials may be particularly useful in optimizing perpendicular magnetic anisotropy (PMA)STT-MRAM devices. For example, a thin CoFeB layer exchange-coupled with a magnetically hard layer (e.g. Co/Pd multilayers) has been widely used for reference layers [17]. When the hard layer has a different crystal asymmetry from the preferred bcc(001) texture in the CoFeB layer, a spacer layer can be inserted between them to prevent the crystallographic template effect and/or atomic interdiffusion from the hard layer. Currently the insertion of a Ta layer between the magnetic hard layer and CoFeB is being pursued [18]. However the possible thickness of the inserted Ta layer is limited since strong ferromagnetic coupling is only observed up to 1-2 atomic monolayers [19]. An appropriate amorphous spacer layer should enable the CoFeB layer to crystallize in the proper bcc(001) structure and thereby enhance tunneling magnetoresistance across a MgO tunnel barrier. It should also be able to act as a diffusion barrier during annealing and be designed to limit dead layers when in proximity to CoFeB. While weak AF-coupling has been previously observed in non-magnetic amorphous metallic spacer layers [20], we show that much stronger ferromagnetic coupling can be obtained using amorphous Fe-Ta alloys enabling the use of a thicker spacerlayer. A similar approach has been used in perpendicular recording media where crystalline Co-Ru alloy spacer layers allow the ferromagnetic coupling to be tuned [12].

## II. Experimental details

To study the exchange coupling through Fe-Ta alloy spacers we used a model system consisting of a magnetically soft PMA layer coupled to a PMA SAF structure (shown schematically in Fig. 1). The samples were grown at room temperature under high vacuum (5E-9 Torr) by DC magnetron sputtering on $SiO_X$-coated Si substrates. The SAF consisted of two Co/Pd multilayers separated by an Ir spacer, $[Pd(0.7)/Co(0.25)]_{x6}/Ir(0.425)/[Co(0.25)/Pd(0.7)]_{x5}/Co(0.25)$ (thickness in nm). The soft layer is a $[Co(0.25)/Pd(0.7)]_{x3}$ multilayer. The softlayer and SAF were separated by Fe-Ta alloy spacers that were grown by co-sputtering from elemental targets. Ta(4)/Pd(4)was used for a seed layer and Pd(4)/Ta(4)was used for a capping layer. Furthermore a series of 20-or 200-nm thick $Fe_x$-$Ta_{1-x}$ alloys (x= 0.5, 0.6, 0.7, 0.75) were grown with 4-nm thick Ta seed and capping layers to measure the susceptibility of the Fe-Ta alloys. Finally, a series of [Fe$_x$Ta$_{1-x}$/CoFeB] multilayers were deposited to study the effect of Fe-Ta alloys on the magnetic dead layer at the Fe$_x$Ta$_{1-x}$-CoFeB interfaces. All the samples were post-annealed under vacuum at 300°C for one hour. Magneto-optical Kerr effect (MOKE) and vibrating sample magnetometry (VSM) were used to characterize the magnetic properties of the samples. The susceptibility of Fe-Ta alloys was obtained from fitting of magnetization hysteresis loops at relatively low fields (300-1000 Oe).

III. Results and discussions

Figure 1a shows the out-of-plane major and minor hysteresis loops for the Co/Pd SAF-based model system with different Ta-spacer thicknesses at room temperature. The coupling strength across the Ta spacer was determined from the shift of the minor loop of the soft layer (inset of Fig. 1a). The coupling field decreases exponentially with increasing Ta thickness (Fig. 1b) and decoupling of the softlayer from the SAF is observed for Ta-thicknesses slightly above one atomic monolayer. The strong reduction of exchange coupling through 1-2 atomic monolayers of Ta is consistent with previous observations [19].

Using Fe$_x$Ta$_{1-x}$ alloys instead of pure Ta allows tuning of the coupling field and critical thicknesses for decoupling (Fig.1b). Both increase with increasing Fe content. We fit the coupling field data at room temperature for each Fe-Ta composition to an exponential form:

$$H_{ex} = H_0 \exp[-\gamma \cdot t] \qquad (1)$$

where $\gamma$ is the decay of the exchange field with thickness $t$ and $H_0$ represents the coupling strength in the hypothetical case of an insertion layer with zero thickness. We find that $\gamma$ decreases with increasing Fe content $x$ and is directly linked to the susceptibility $\chi$ of the Fe-Ta alloys (Fig.1c). From Fig.1d, one can see that higher susceptibility leads to stronger exchange coupling for comparable spacer thicknesses and thus to smaller decay rates of the exchange field.

To further explore the correlation of exchange coupling with susceptibility we measured hysteresisloops at different temperatures for 2.1 and 2.6-nm $Fe_{75}Ta_{25}$ spacers and for a 2.0-nm $Fe_{60}Ta_{40}$ spacer. Additionally the susceptibility as a function of temperature for a 20-nm thick $Fe_{75}Ta_{25}$ layer and for a 200-nm thick $Fe_{60}Ta_{40}$ layer (Fig.2 a-b) was measured. Both $\chi$ and coupling field increase with decreasing temperature. If $1/\chi$ is plotted as a function of temperature a Curie-Weiss behavior is observed (Fig.2c). The deduced Curie Temperature $T_C$ for $Fe_{75}Ta_{25}$ is 75K, whereas it is -20K for $Fe_{60}Ta_{40}$, meaning that ferromagnetic order cannot be obtained for a $Fe_{60}Ta_{40}$ alloy even at low temperature. In Fig.2d the exchange field is plotted for all three samples as a function of susceptibility (tuned via the temperature). Increasing susceptibility leads to enhanced exchange coupling as expected. The coupling strength scales roughly with the logarithm of the susceptibility. We do not have a quantitative model relating the coupling strength and susceptibility, however previous calculations of ferromagnetic/paramagnetic multilayers showed that the induced magnetic order in the paramagnetic layer falls off quickly from the interface and is strongly temperature dependent, in agreement with our findings [21].

As discussed above, one application for an amorphous coupling layer is to couple a high-anisotropy magnetic layer to a CoFeB layer to act as a reference layer for PMA STT-MRAM. Figure 3 shows the magnetic moment of CoFeB in CoFeB/spacer (Ta or Fe-Ta alloy) multilayers as a function of the nominal CoFeB layer thicknesses. The moment increases linearly with nominal CoFeB thickness in all the samples. The minimum thickness of CoFeB which is needed in order to observe magnetic moment in the case of Ta is commonly interpreted as a magnetic dead layer at the CoFeB/Ta interfaces [22]. The magnetic dead layer thickness is extracted from the x-intercept values. For a pure Ta spacer, the total magnetic dead layer for each CoFeB layer sandwiched by Ta was 1.3 nm (i.e. 0.65 nm per CoFeB-Ta interface), which persists even at 75 K. We find the magnetic dead layer decreases with increasing Fe content in the Fe-Ta layers and disappears for an Fe content of 75at.% at room temperature (Fig.3a). For higher Fe content and low temperature we observe a negative x-intercept. This is interpreted as an induced moment in the Fe-Ta spacer by CoFeB at low temperatures which is often seen in Co/Pd and Co/Pt multilayers [23]. In Fig. 3c this induced moment for $Fe_{75}Ta_{25}$ is expressed in units of CoFeB thickness similar to a magnetic dead layer. In Fig. 3d the anisotropy field of CoFeB/$Fe_xTa_{1-x}$ that is estimated from the saturation of the magnetization in an out-of-plane magnetic field is shown. The values are negative indicating that that anisotropy is dominated by shape anisotropy. However, the effective shape anisotropy is slightly lower for samples with equal moment per area ratio m/S when the Fe content in Fe-Ta is high. In a simple model for the anisotropy of those [CoFeB/FeTa] multilayers considering shape anisotropy as $K_D$ and a perpendicular interface anisotropy of $K_S$ per CoFeB/FeTa interface we obtain:

$$H_K = \frac{2K_S \cdot 10}{m/S} + \frac{2K_D}{M} \qquad (2)$$

If we apply this model to our data we find $K_S$ around 0.1 erg/cm$^2$ per Ta/CoFeB and per CoFeB/$Fe_{50}Ta_{50}$ interface and about 0.2 erg/cm$^2$ per CoFeB/$Fe_{75}Ta_{25}$ interface.

IV. Conclusion

We have shown that the magnetic exchange coupling through amorphous Fe-Ta alloys can be tuned via the magnetic susceptibility which depends on Fe content and the temperature. Such alloys are of high interest for engineering magnetic devices such as PMA tunnel junctions since they allow ferromagnetic coupling of magnetic layers with differing crystalline symmetry. Furthermore these layers suppress magnetic dead layers and also act as anti-diffusion barrier between hard magnetic materials and the CoFeB/MgO system during subsequent annealing processes. These properties make Fe-Ta alloy spacers superior to Ta spacers since magnetic coupling can be conserved for thicknesses which are at least 5-6 times larger than in the case of pure Ta. While we studied amorphous and paramagnetic Fe-Ta alloys, there is a large set of amorphous binary Fe, Co and Ni alloys with transition metals such as Ta, W, Zr etc. for which the onset of magnetic order depends on the alloy concentration and temperature [24-26]. This allows choosing chemically adapted alloys to block interdiffusion between specific materials while coupling them magnetically.

Acknowledgement

Authors thank Richard Choi and Dr. Fred Spada for assistance with their MOKE set-ups.

Research at UCSD was supported by NSF Award # DMR-1008654and by UC Discovery Grant 21294.

**References:**
[1]     M. N. Baibich, et al., Phys. Rev. Lett. 61, 2472 (1988).
[2]     G. Binasch, et al., Physical Review B 39, 4828 (1989).
[3]     H.-C.Tong, F. Liu, K. Stoev,Y. Chen, X.Shi, C.Qian; Jour. Mag. Mag. Mat. 239 106-111 (2002)
[4]     J.-M. Hu, Z. Li, L.-Q. Chen, and C.-W. Nan; Nat. Comm. 2,553 (2011)
[5]     B. N. Engel, J. Åkerman, B. Butcher, R.W. Dave, M. De Herrera, M. Durlam, G. Grynkewich, J. Janesky, S.V. Pietambaram, N.D.Rizzo, J.M.Slaughter, K.Smith, J.J. Sun,and S.Tehrani, IEEE Trans. Mag. 41, 132 (2005).
[6]     E. E. Fullerton, D.T. Margulies, M.E.Schabes, M. Carey,B. Gurney, A. Moser, M. Best, G. Zeltzer, K. Rubin, and H. Rosen., Appl. Phys. Lett.77, 3806 (2000).
[7]     E. N. Abarra, A. Inomata, H. Sato, I.Okamoto, and Y. Mizoshita, Appl. Phys. Lett. 77, 2581 (2000).

[8] A. Moser, K.Takano, D. Margulies, M. Albrecht, Y. Sonobe, Y. Ikeda, S. Sun and E. E. Fullerton, Jour. of Phys. D-Applied Physics 35, R157 (2002).

[9] S. C. Byeon, A. Misra, and W. D. Doyle; IEEE Trans. Mag. 40, 4, 2386-2388 (2004)

[10] A. Berger, N.Supper, Y. Ikeda, B. Lengsfield, A. Moser, and E.E.Fullerton; Appl. Phys. Lett. 93, 122502 (2008).

[11] R. H. Victora and X. Shen, IEEE Transactions on Magnetics 41, 2828 (2005).

[12] N. F. Supper, D.T. Margulies, A. Moser, A. Berger, H. Do, and E. E. Fullerton, IEEE Trans. Mag. 41, 3238 (2005).

[13] I. Yulaev, M.V. Lubarda, S. Mangin, V.Lomakin, and E. E. Fullerton, Appl. Phys. Lett.99, 132502 (2011).

[14] B. K. Kuanr, Y.V. Khivintsev, A. Hutchison, R.E. Camley, and Z.J. Celinski, IEEE Transactions on Magnetics 43, 2648 (2007).

[15] W. H. Rippard, M. R. Pufall, and T. J. Silva, Applied Physics Letters 82, 1260 (2003).

[16] C.L. Chien, S.H. Liou, B. K. Ha, and K. M. Unruh, J. Appl. Phys. 57(1), 3539 (1985)

[17] M. T.Rahman, A. Lyle, G. Hu, W. J. Gallagher, and J. -P. Wang; J. Appl. Phys. 109, 07C709 (2011)

[18] D. C. Worledge, G. Hu, D. W. Abraham, J. Z. Sun, P. L. Trouilloud, J. Nowak, S. Brown, M. C. Gaidis, E. J. O'Sullivan, and R. P. Robertazzi; Appl. Phys. Lett. 98, 022501 (2011)

[19] V. Sokalski, M.T. Moneck, E.Yang, and J.-G.Zhu; Appl. Phys. Lett. 101, 072411 (2012)

[20] D.E. Bürgler, D.M. Schaller, C.M. Schmidt, F. Meisinger, J. Kroha, J. McCord, A. Hubert, and H.-J. Güntherodt;Phys. Rev. Lett.80, 4983 (1998).

[21] R.W. Wang and D. L. Mills; Phys. Rev. B 46, 18 11681 (1992)

[22] S. Y. Jang, S.H. Lim, andS.R.Lee;J. Appl. Phys. 107, 09C707 (2010)

[23] B. N. Engel, C. D. England, R. A. Vanleeuwen, M. H. Wiedmann, and C. M. Falco, Phys. Rev. Lett, 67, 1910 (1991)

[24] T. Egami and Y. Waseda; Journal of Non-Crystalline Solids 64;113-134 (1984)

[25] M. Naoe, H. Kazama, Y. Hoshi, and S. Yamanaka; J. Appl. Phys. 53(11), 7846 (1982)

[26] M. Sostarich; J. Appl.Phys. 69 (8), 6141, (1991)

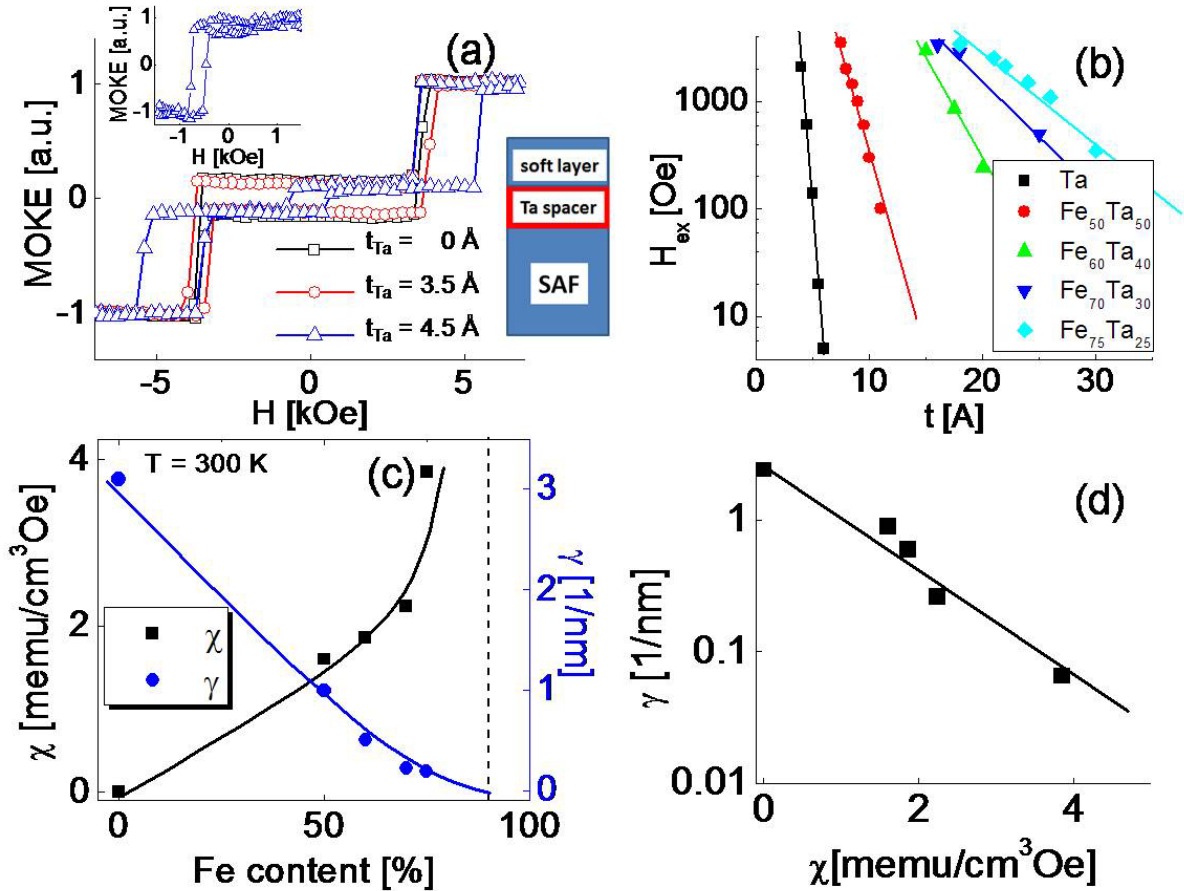

**Figure 1:** Exchange coupling through Fe-Ta layers studied in a model system consisting of a [Co/Pd] /Ir/[Co/Pd]-based SAF layer and a [Co/Pd] based soft layer at room temperature: Ta(4)/Pd(4)/[Pd0.7/Co0.25]x6/Ir(0.425)/[Co0.25/Pd0.7]x5/Co(0.25)/Fe-Ta spacer/[Co0.25/Pd0.7]$_{x3}$/Pd(4)/Ta(4). (a) Hysteresis major loops measured by magneto-optical Kerr effect (MOKE) for different Ta thicknesses $t_{Ta}$ and minor loop for $t_{Ta}$ = 4.5Å (inset). (b) Exchange field $H_{ex}$ between SAF and soft layer for different $Ta_{1-x}Fe_x$ spacer compositions as a function of thickness at room temperature. Lines are fits to Eq. (1). (c) Decay rate $\gamma$ of the exchange field $H_{ex}$ through $Ta_{1-x}Fe_x$ spacers as a function of the Fe content $x$. Decay-rates $\gamma$ are obtained from fits to data in Fig. 1(b). Susceptibility $\chi$ is obtained from in-plane VSM hysteresis loops for 20-nm-thick $Fe_xTa_{1-x}$ layers. Lines are guides to the eye. (d) Decay rate $\gamma$ as a function of the susceptibility $\chi$.

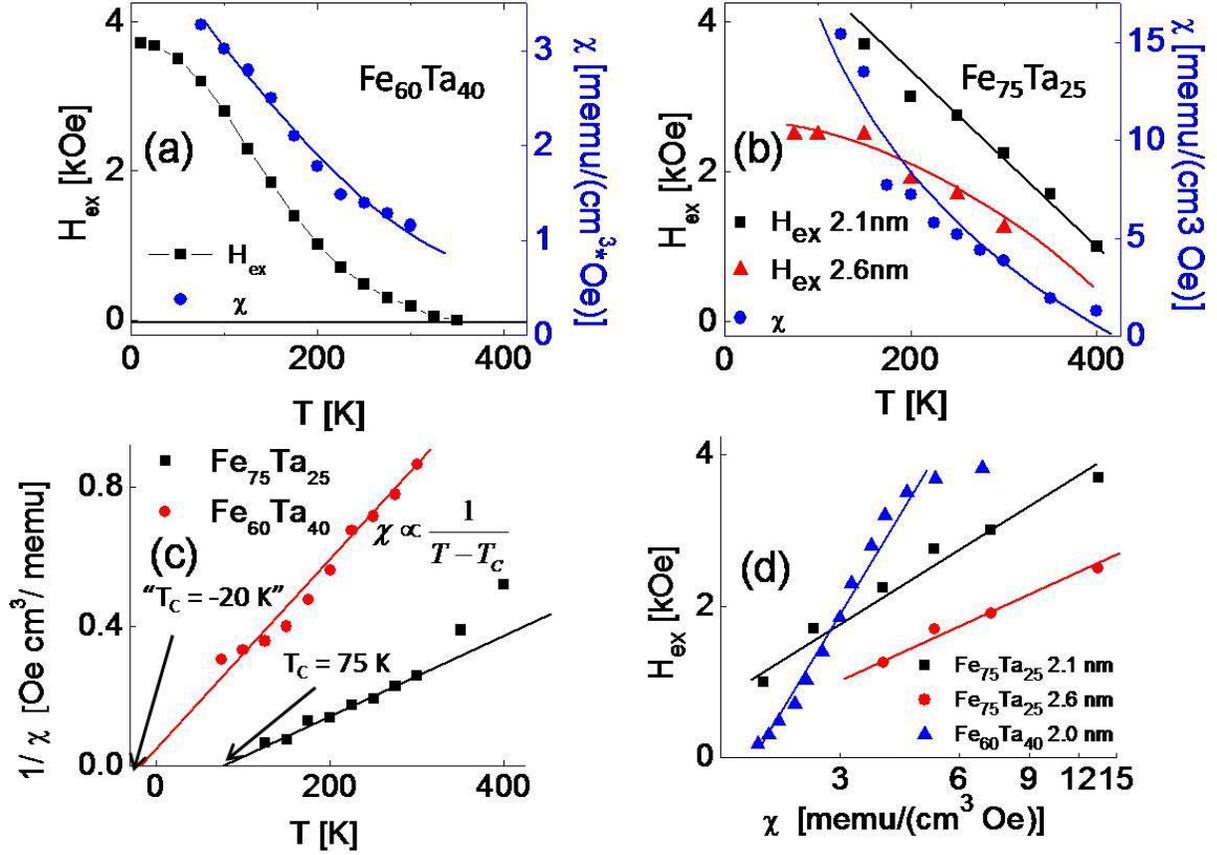

Figure 2: Temperature dependence of susceptibility χ and coupling field $H_{ex}$ for FeTa spacers (a) $H_{ex}$ between the SAF and soft layers *vs.* temperature T through a 2-nm-$Ta_{40}Fe_{60}$ spacer and paramagnetic susceptibility χ vs. temperature for a $Ta_{40}Fe_{60}$ alloy. (b) $H_{ex}$ between the SAF and soft layers *vs.* T through a 2.1-nm and a 2.6-nm thick $Ta_{25}Fe_{75}$ spacer and paramagnetic susceptibility χ vs. temperature for a $Ta_{25}Fe_{75}$ alloy. (c) Reciprocal susceptibility 1/χ as a function of T for $Ta_{40}Fe_{60}$ and $Ta_{25}Fe_{75}$ alloy films. (d) Exchange field $H_{ex}$ between the SAF and soft layers for a 2.1-nm and 2.6-nm $Ta_{25}Fe_{75}$ spacers and a 2.0-nm $Ta_{40}Fe_{60}$ spacer as a function of the susceptibility χ.

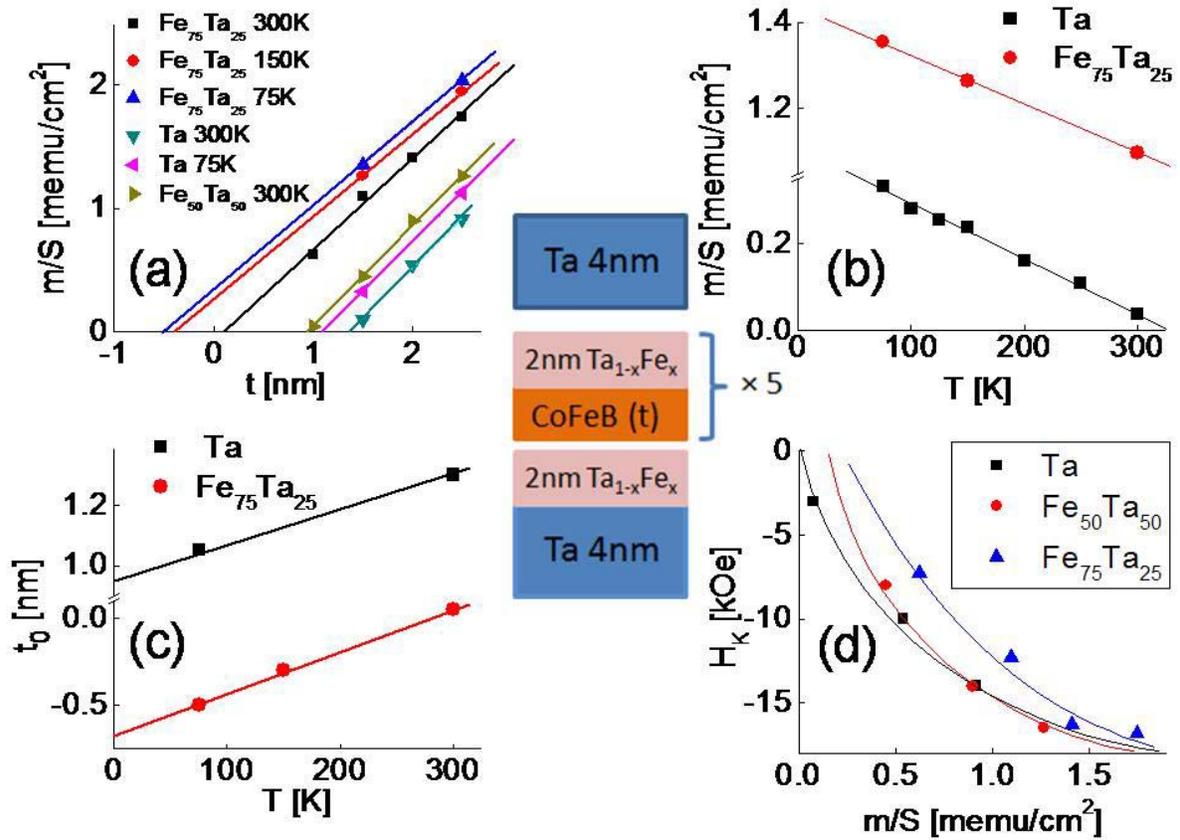

Figure 3: Magnetic properties of $[Co_{20}Fe_{60}B_{20}/(Fe)Ta]$ multilayers for different $Ta_{1-x}Fe_x$ alloys in a $[Ta_{1-x}Fe_x(2nm)/CoFeB\ (t\ nm)]x5/Ta_{1-x}Fe_x(2nm)$ multilayer. (a) Magnetic moment per area for CoFeB/$Ta_{1-x}Fe_x$ multilayers as a function of the CoFeB thickness for different T. (b) Magnetic moment per area for CoFeB (1.5 nm)/$Ta_{1-x}Fe_x$ multilayers as a function of the T. (c) Difference between nominal and effective CoFeB thickness $t_0$ as a function of temperature. (d) Out of plane anisotropy field $H_K$ vs. moment per area $m/S$ for CoFeB/$Ta_{1-x}Fe_x$ multilayers at room temperature.

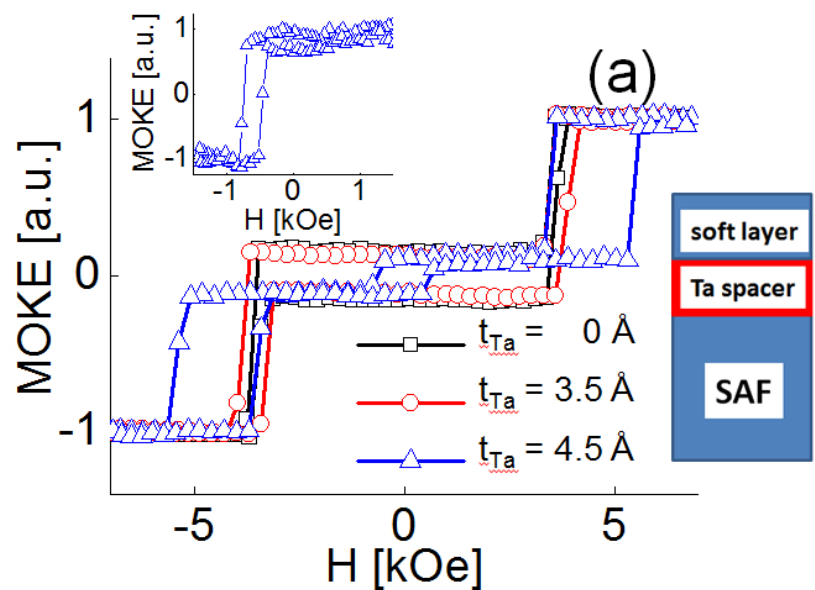
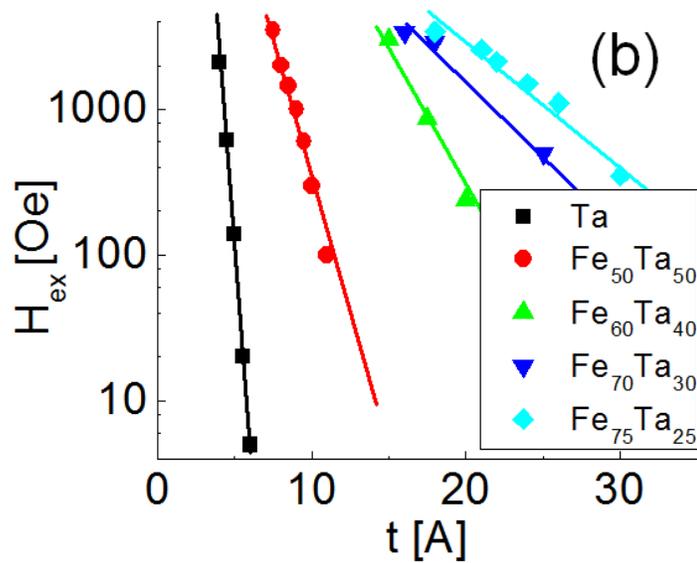
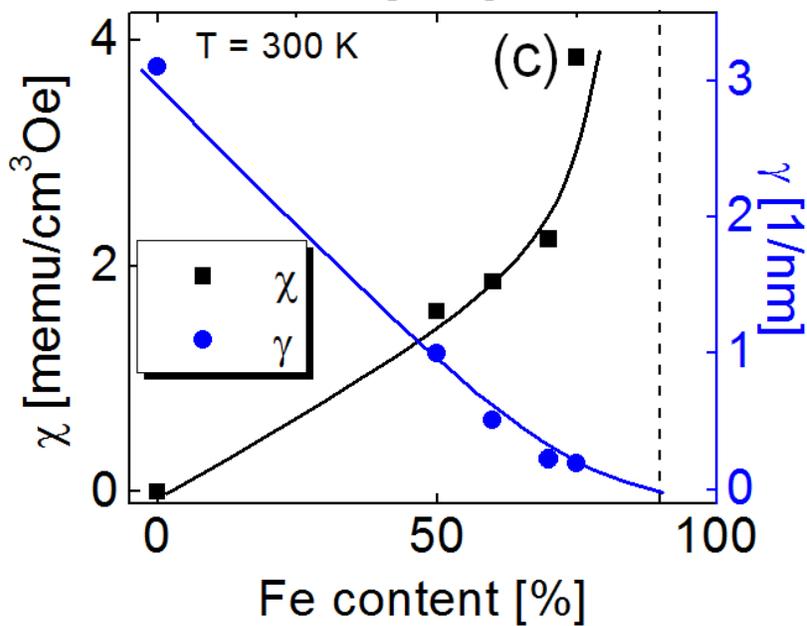
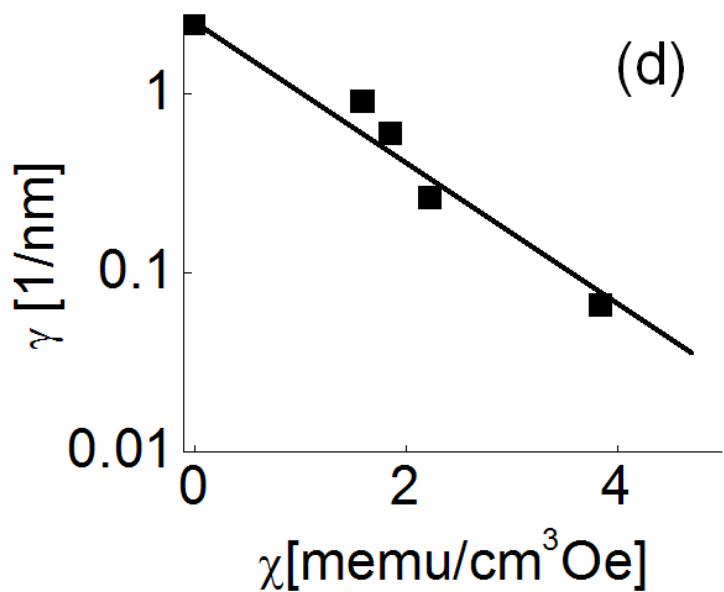

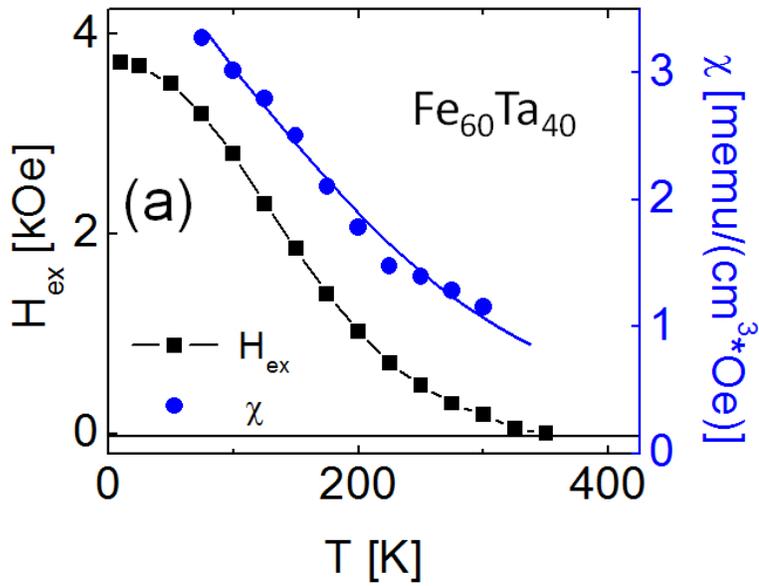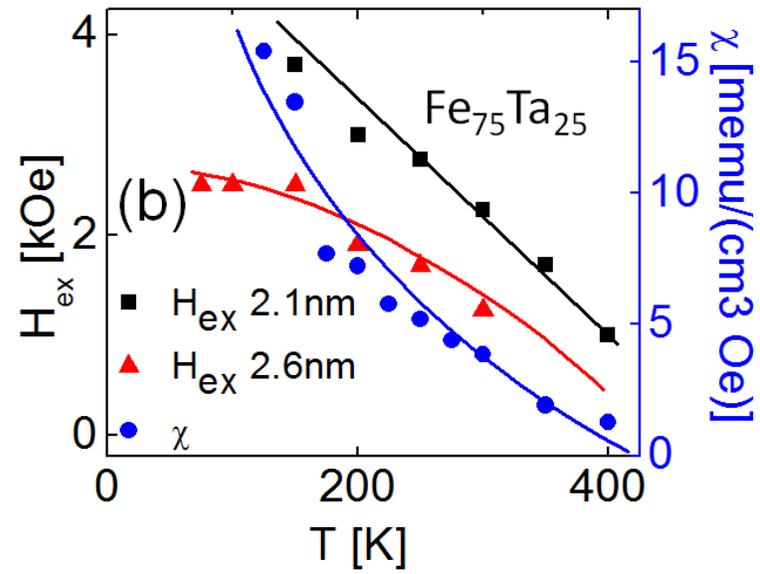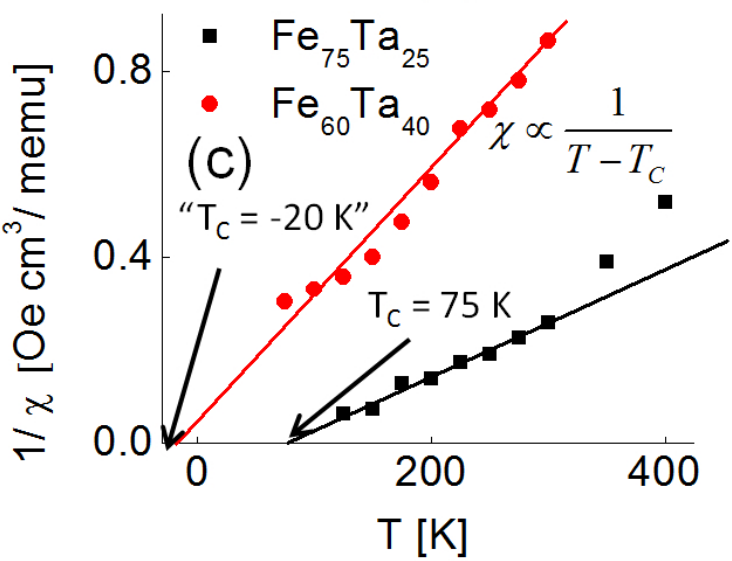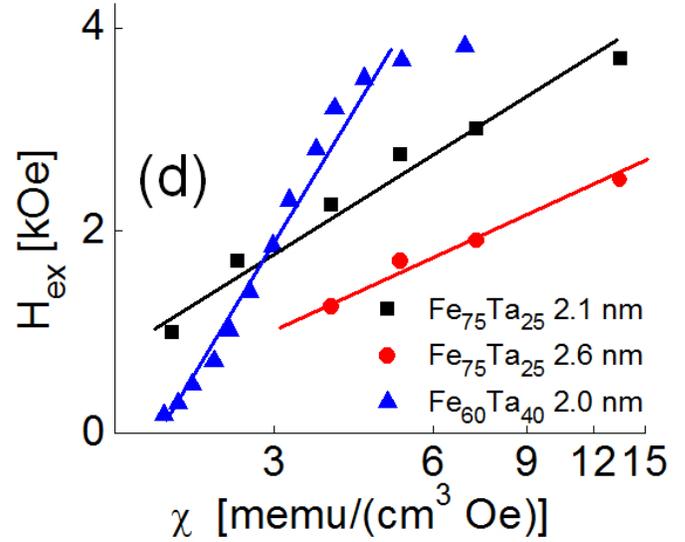

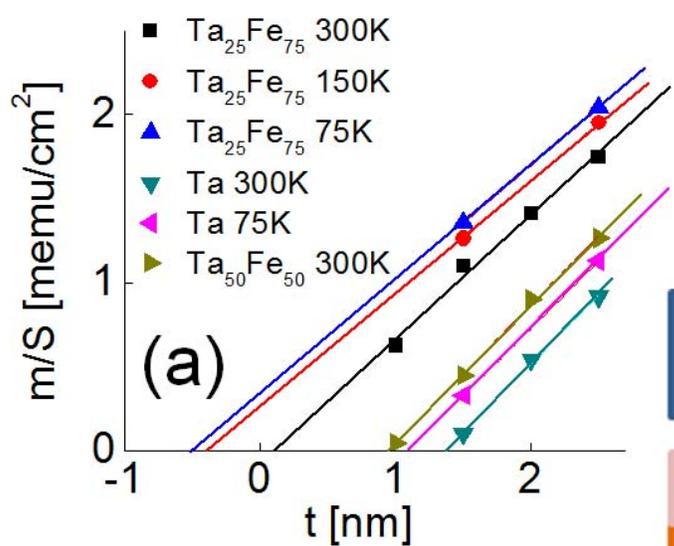
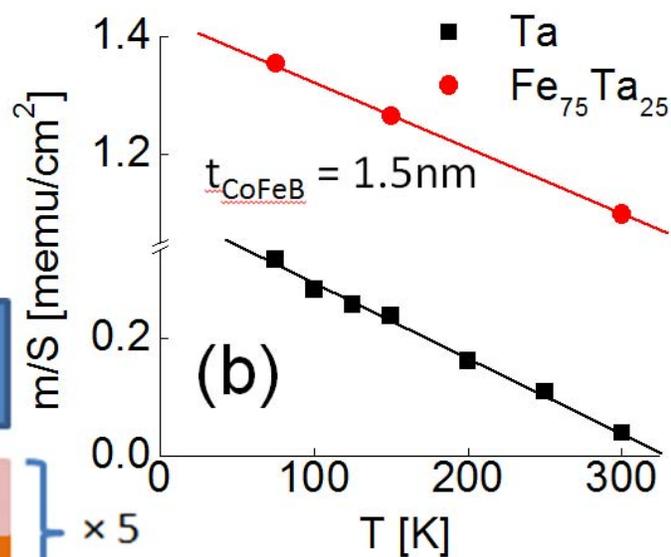
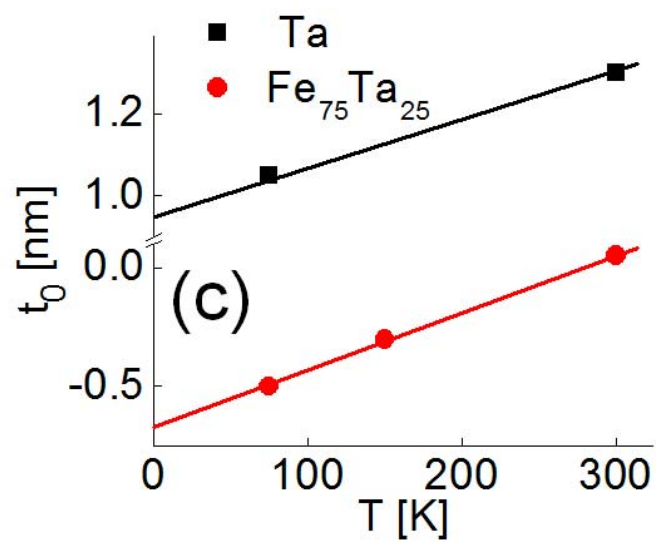
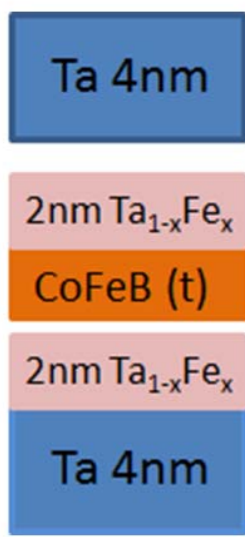
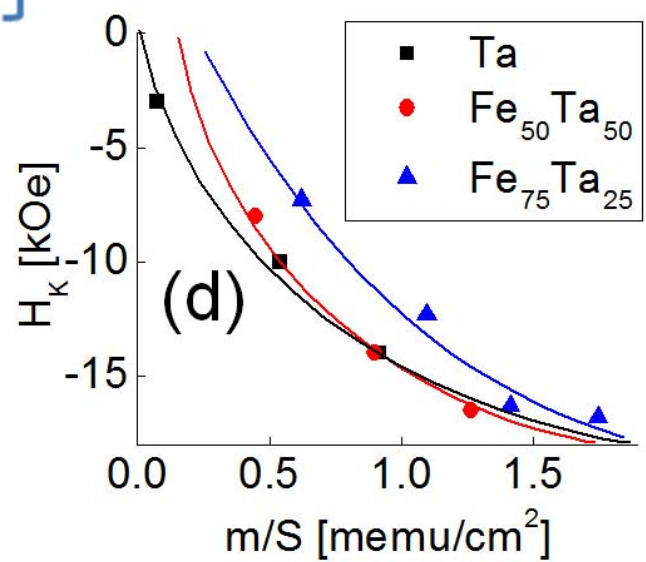